\documentclass[aoas,nameyear,dvips]{arximspdf}
\usepackage{graphics}
\doi{10.1214/09-AOAS320}
\volume{4}
\issue{3}
\pubyear{2010}
\firstpage{1311}
\lastpage{1341}

\begin{document}
\begin{frontmatter}

\title{Approaches for multi-step density forecasts with application to aggregated wind power}
\pdftitle{Approaches for multi-step density forecasts with application to aggregated wind power}
\runtitle{Multi-step wind power density forecasts}

\begin{aug}
\author[A]{\fnms{Ada} \snm{Lau}\ead[label=e1]{ada.lau@oxford-man.ox.ac.uk}\corref{}\thanksref{aut1}}
and
\author[B]{\fnms{Patrick} \snm{McSharry}\ead[label=e2]{patrick@mcsharry.net}\thanksref{aut2}}
\thankstext{aut1}{Supported by the Oxford--Man Institute of Quantitative Finance.}
\thankstext{aut2}{Supported by a Royal
Academy of Engineering/EPSRC Research
Fellowship and by the European
Commission under the SafeWind Project
(ENK7-CT2008-213740).}
\runauthor{A. Lau and P. M{\normalfont \textsc{c}}Sharry}
\affiliation{University of Oxford}
\address[A]{Oxford--Man Institute \\
University of Oxford\\
Eagle House, Walton Well Road \\
Oxford, OX2 6ED\\
United Kingdom \\
\printead{e1}} 
\address[B]{Smith School of Enterprise\\
\quad and the Environment\\
University of Oxford\\
Hayes House\\
75 George Street\\
Oxford, OX2 2BQ\\
United Kingdom \\
\printead{e2}}
\end{aug}

\received{\smonth{9} \syear{2009}}
\revised{\smonth{12} \syear{2009}}

\begin{abstract}
The generation of multi-step density forecasts for non-Gaussian data
mostly relies on Monte Carlo simulations which are computationally
intensive. Using aggregated wind power in Ireland, we study two
approaches of multi-step density forecasts which can be obtained from
simple iterations so that intensive computations are avoided. In the
first approach, we apply a logistic transformation to normalize the
data approximately and describe the transformed data using ARIMA--GARCH
models so that multi-step forecasts can be iterated easily. In the
second approach, we describe the forecast densities by truncated normal
distributions which are governed by two parameters, namely, the
conditional mean and conditional variance. We apply exponential
smoothing methods to forecast the two parameters simultaneously. Since
the underlying model of exponential smoothing is Gaussian, we are able
to obtain multi-step forecasts of the parameters by simple iterations
and thus generate forecast densities as truncated normal distributions.
We generate forecasts for wind power from 15 minutes to 24 hours ahead.
Results show that the first approach generates superior forecasts and
slightly outperforms the second approach under various proper scores.
Nevertheless, the second approach is computationally more efficient and
gives more robust results under different lengths of training data. It
also provides an attractive alternative approach since one is allowed
to choose a particular parametric density for the forecasts, and is
valuable when there are no obvious transformations to normalize the
data.
\end{abstract}


\end{frontmatter}

\setcounter{footnote}{2}

\section{Introduction}\label{sec:Intro}

Wind power forecasts are essential for the efficient operation and
integration of wind power into the national grid. Since wind is
variable and wind energy cannot be stored efficiently, there are
risks of power shortages during periods of low wind speed. Wind
turbines may also need to be shut down when wind speeds are too high,
leading to an abrupt drop of power supply. It is extremely important
for power system operators to quantify the uncertainties of wind power
generation in order to plan for system reserve efficiently
[\citet{Doherty2005}]. In addition, wind farm operators require
accurate estimations of the uncertainties of wind power generation to
reduce penalties and maximize revenues from the electricity market
[\citet{Pinson2007}].

Since the work of \citet{Brown1984} in wind speed forecasting using
autoregressive models, there has been an increasing amount of research
in wind speed and wind power forecasts. Most of the early literature
focuses on point forecasts, and in recent years more emphasis
has been placed on probabilistic or density forecasts because of the need to
quantify uncertainties. However, the number of studies on multi-step
density forecasts is still relatively small, not to mention the
evaluation of forecast performances for horizons $h>1$. Early works on
multi-step density forecasts can be found in \citet{Davies1988} and
\citet{Moeanaddin1990}, where the densities are estimated using
recursive numerical quadrature that requires significant computational
time. \citet{Manzan2008} propose a nonparametric way to generate density
forecasts for the U.S. Industrial Production series, which is based on
bootstrap methods. However, Monte Carlo simulations are required and
this approach is also computationally intensive.

One of the approaches to wind power forecasting is to focus on the
modeling of wind speed and then transform the data into wind power
through a power curve [\citet{Sanchez2006}]. An advantage is that
wind speed time series are smoother and more easily described by linear
models. However, a major difficulty is that the shape of the power
curve may vary with time, and also it is difficult to quantify the
uncertainties in calibrating the nonlinear power curve. Another
approach is to transform meteorological forecasts into wind power
forecasts, where ensemble forecasts are generated from sophisticated
numerical weather prediction (NWP) models [\citet{Taylor2006}, \citet{Pinson2009}]. This approach is able to produce reliable wind power
forecasts up to 10 days ahead, but it requires the computation of a
large number of scenarios as well as expensive NWP models. A third approach to wind power forecasting focuses
on the direct statistical modeling of wind power time series. In this
case the difficulty lies on the fact that wind power time series are
highly nonlinear and non-Gaussian. In particular, wind power time
series at individual wind farms always contain long chains of zeros and
sudden jumps from maximum capacity to a low value due to gusts of wind
since turbines have to be shut down temporarily. Nevertheless, it has
been shown that statistical time series models may outperform
sophisticated meteorological forecasts for short forecast horizons
within 6 hours [\citet{Milligan2003}]. Extensive reviews of the short
term state-of-the-art wind power prediction are contained in
\citet{Landberg2003}, \citet{Giebel2003} and
\citet{Costa2008}, in which power curve models, NWP models and
other statistical models are discussed.

In this paper we adopt the third approach and consider modeling the
wind power data directly. We aim at short forecast horizons within 24
hours ahead, since for longer forecast horizons the NWP models may be
more reliable. As mentioned above, wind power time series are highly
nonlinear. Aggregating the individual wind power time series will
smooth out the irregularities, resulting in a time series which is more
appropriately described by linear models under suitable
transformations. Aggregated wind power generation is also more relevant
to power companies since they mainly consider the total level of wind
power generation available for dispatch. Thus, it is economically
important to generate reliable density forecasts for aggregated wind
power generation.

For this reason, as a first study, this paper considers the modeling of
aggregated wind power time series. One may argue that utilizing
spatiotemporal correlations among individual wind farms may improve the
results in forecasting aggregated wind power. We will show in Section
\ref{sec:Evaluation} that this is not the case here, at least by the
use of a simple multiple time series approach. Unless one is interested
in the power generated at individual wind farms, it is more appropriate
to forecast the aggregated wind power as a univariate time series. We
propose two approaches of generating multi-step ahead density forecasts
for wind power generation, and we demonstrate the value of our
approaches using wind power generation from 64 wind farms in Ireland.
In the first approach, we demonstrate that the logistic function is a
suitable transformation to normalize the aggregated wind power data. In
the second approach, we describe the forecast densities by truncated
normal distributions which are governed by two parameters, namely, the
conditional mean and conditional variance. We apply exponential
smoothing methods to forecast the two parameters simultaneously. Since
the underlying model of exponential smoothing is Gaussian, we are able
to obtain multi-step forecasts of the parameters by simple iterations
and thus generate forecast densities as truncated normal distributions.
Although the second approach performs
similarly to the first in terms of our
evaluation of the wind power forecasts, it
has numerous advantages. It is
computationally more efficient, its forecast performances are more
robust, and it provides the flexibility to choose a suitable parametric
function for the density forecasts. It is also valuable when there are
no obvious transformations to normalize the data.

Our paper is organized as follows. In Section \ref{sec:data} we
describe the wind power data that we use in our study. Then we explain
the two approaches of generating multi-step density forecasts in
Section \ref{sec:Approach}. The first approach concerning the logistic
transformation is described in Section \ref{sec:model}, while in
Section \ref{sec:ESTrunNorm} we give the details on the second approach
using exponential smoothing methods and truncated normal distributions.
In Section \ref{sec:Evaluation} we construct 4 benchmarks to gauge the
performances of our approaches, and we evaluate the forecast
performances using various proper scores. Finally, we conclude our
paper in Section \ref{sec:Conclusion}, where we summarize the benefits
of our approaches and discuss important future research directions.

\section{Wind power data}\label{sec:data}

\begin{figure}

\includegraphics{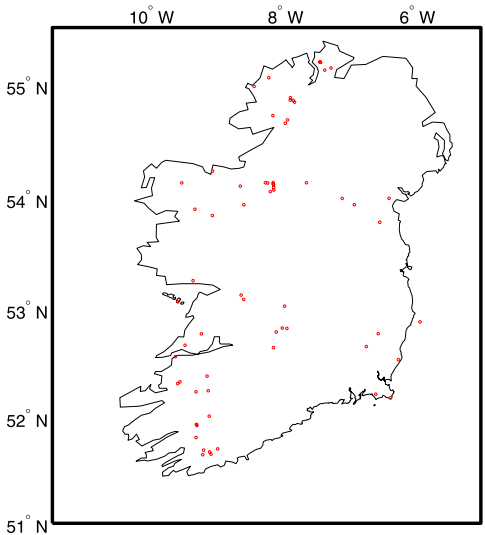}

\caption{The locations of 64 wind farms in Ireland. There are 68 wind farms and wind power
time series in the raw data, but 4 pairs of
wind farms are so close that they are
essentially extensions from the
corresponding old wind farm. As a result,
we simply consider 64 wind farms here. The wind farms are distributed throughout
Ireland, and Arklow Banks is the only
offshore wind farm.}\label{fig:FarmLocation}
\end{figure}%
We consider aggregated wind power generated from 64 wind farms in
Ireland for approximately six months from 13-Jul-2007 to 01-Jan-2008.
The data are recorded every 15 minutes, giving a total number of 16,512
observations during the period. The locations of the wind farms
are
shown in Figure~\ref{fig:FarmLocation}. One of the wind farms, known as Arklow
Banks, is offshore.\footnote{Detailed information of
individual wind farms, such as latitude, longitude and capacity, is
provided by Eirgrid plc and can be found in \citet{Lau2010}.} We
sum up the capacities\footnote{The capacity is the maximum output of a
wind farm when all turbines operate at their maximum nominal power.} of
all wind farms and the total capacity is 792.355~MW. In order to
facilitate comparisons between data sets with different capacities, we
normalize the aggregated wind power by dividing by the total capacity,
that is, 792.355~MW, and so the normalized data is bounded within $[0,1]$.
We have checked that forecast results, in particular, for approaches
involving nonlinear transformations, are in fact insensitive to the
exact value of normalization.\footnote{In our paper the value of
normalization must not be smaller than the total capacity since we will
consider the logistic transformation (\ref{eq:logit}).} We dissect the
data into a training set of about 4 months (the first 11,008 data
\begin{figure}

\includegraphics{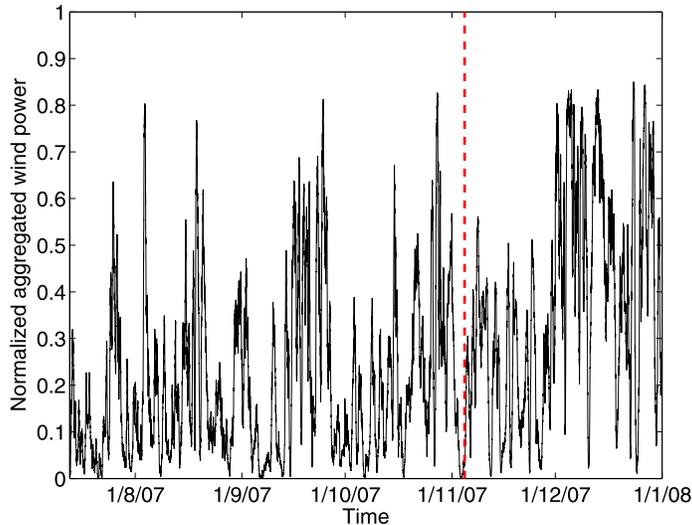}

\caption{Time series of normalized aggregated wind power from 64 wind
farms in Ireland, where the aggregated wind power is normalized by the
total capacity of 792.355~MW. The data are dissected into a training set
and a testing set as shown by the dashed line. About four months of
data are used for parameter estimation, and the remaining two months of
data are used for out-of-sample evaluation.}\label{fig:NormWP_bw}
\end{figure}
\begin{figure}

\includegraphics{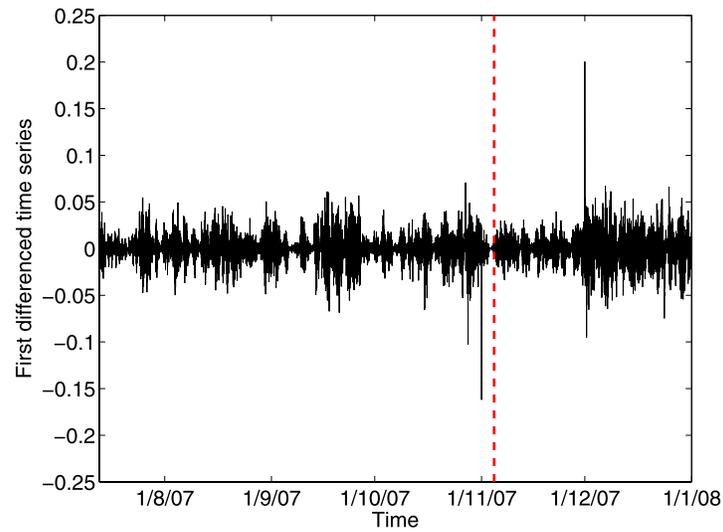}

\caption{First differences of normalized aggregated wind power. It is
clear that the variance changes with time, and there is volatility
clustering as well as sudden spikes. The data are dissected by the
dashed line into a training set and a testing set.}\label{fig:NormWPDiff_bw}
\end{figure}
\begin{figure}

\includegraphics{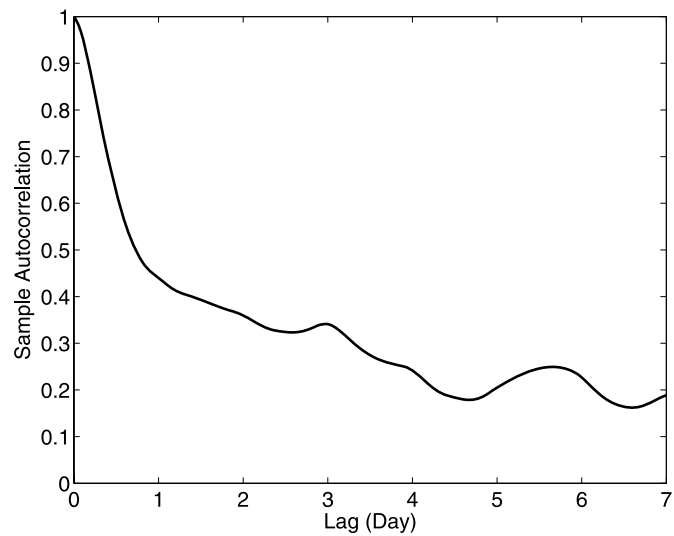}

\caption{Sample ACF of the time series of normalized aggregated wind
power up to a lag of 7 days. The autocorrelations decay very slowly. It
shows that the wind power data are highly correlated and may
incorporate long memory effects.} \label{fig:NormWP_ACF_bw}
\end{figure}
\begin{figure}

\includegraphics{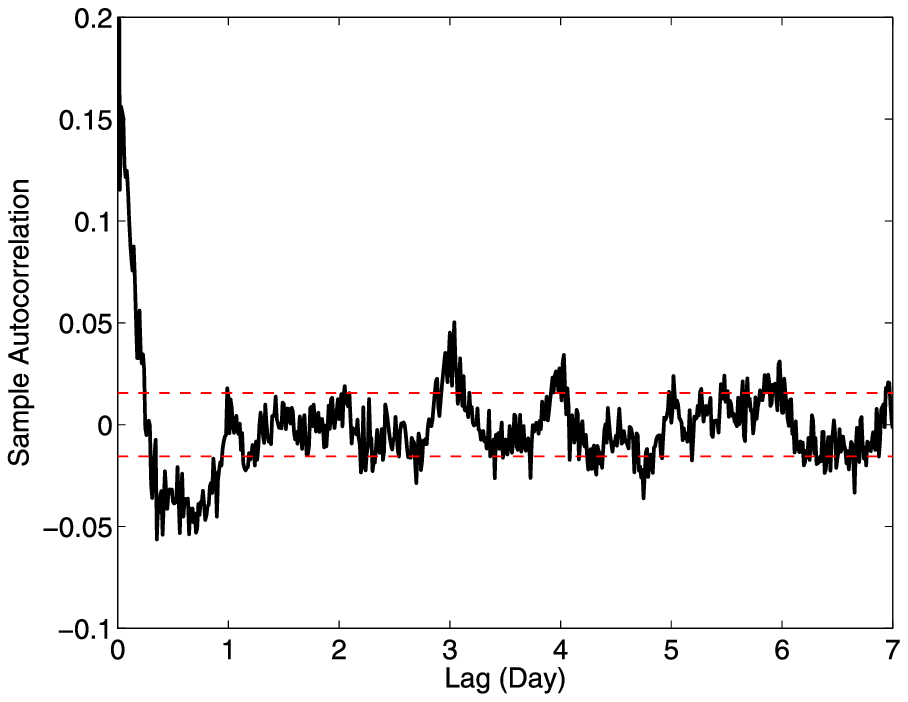}

\caption{Sample ACF of the first differences of normalized aggregated
wind power up to a lag of 7 days. The
dashed lines are the confidence bounds at 2
standard deviations, assuming that the data
follow a Gaussian white noise process. The autocorrelations are
significantly reduced, but they are still significant up to a lag of 7
days.} \label{fig:NormWPDiff_ACF_bw}
\end{figure}%
points) for parameter estimation, and a testing set of about two months
(the remaining 5504 data points) for out-of-sample forecast
evaluations. Figures \ref{fig:NormWP_bw} and \ref{fig:NormWPDiff_bw}
show the original and the first differences of the normalized
aggregated wind power respectively. It is clear that wind power data
are nonstationary. The variance is changing with time, showing clusters of high and low variability. Also, there are some occasional spikes.
Figures~\ref{fig:NormWP_ACF_bw} and \ref{fig:NormWPDiff_ACF_bw} show the
autocorrelation function of the wind power and its first differences
respectively. Autocorrelation is significantly reduced by taking first
differences.

\begin{figure}

\includegraphics{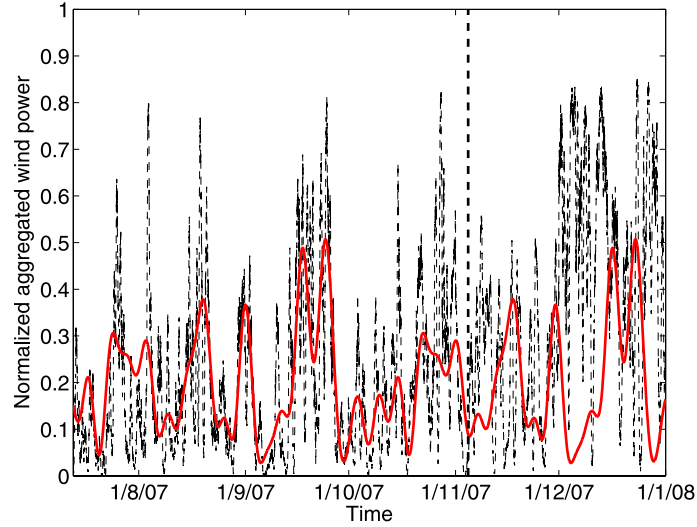}

\caption{Long term seasonality appears in the wind data. We regress the
data in the training set with 16 harmonics of sines and cosines with
periods $T = j/(90 \times 96)$, $j = 1, \ldots, 16$, so that the maximum
period is 90 days. The fit gives an $R^2 = 0.395$. The thin dashed line
is the observed normalized wind power and the solid line is the fitted
time series with a cycle of 90 days. The vertical dashed line dissects
the data into a training set and a testing set.}\label{fig:seasonality}
\end{figure}

Since our aim is to generate short term forecasts up to 24 hours ahead,
we do not focus on modeling any long term seasonality, which often
appears in wind data due to the changing wind patterns throughout the
year. For example, we can model a cycle of 90 days by regressing the
data in the training set with 16 harmonics of sines and cosines with
periods $T = j/(90 \times 96), j = 1, \ldots, 16$. This gives a fitted
time series as shown in Figure \ref{fig:seasonality} with $R^2 = 0.395$.
One may then model the deseasonalized data, but studies
show that results may be worse than those obtained by modeling the
seasonality directly [\citet{Jorgenson1967}]. On the other hand,
we are more interested in the diurnal cycle since it plays a more
important role in intraday forecasts. Diurnal cycles may appear in wind
data due to different temperatures and air pressures during the day and
the night, and wind speeds are sometimes larger during the day when
convection currents are driven by the heating of the sun. Thus, we try
to fit the training data with harmonics of higher frequencies, such as
those with $T = j/96$ where $j$ is an integer. However, results show
that those harmonics cannot help us to explain the variances in the
data, and, thus, we decide to exclude the modeling of any diurnal cycle
in this paper.

\begin{figure}

\includegraphics{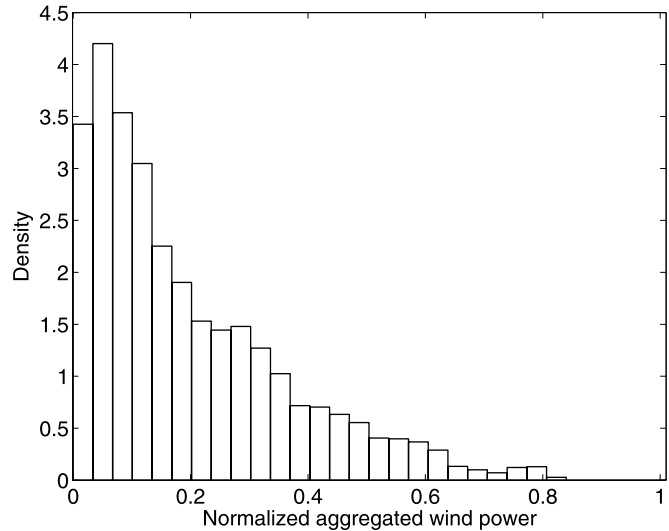}

\caption{Unconditional empirical density of the normalized aggregated
wind power, fitted using the data in the training set. The density is
clearly non-Gaussian since the data is bounded. The density is skewed
and has a sharper peak than the Gaussian distribution. This density
gives the climatology forecast benchmark.}\label{fig:epdf_bar_bw}
\end{figure}

\begin{figure}

\includegraphics{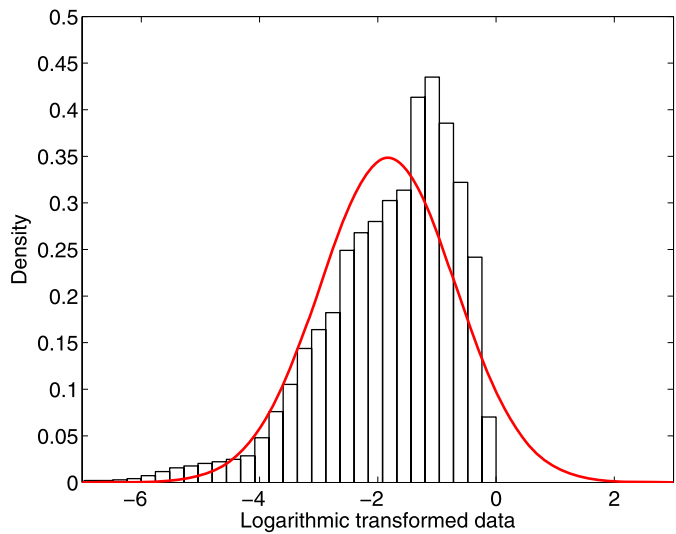}

\caption{Density of the wind power data after applying the logarithmic
transformation, which remains non-Gaussian. The logarithmic
transformation is a common transformation to convert wind speed data
into an approximate Gaussian distribution, but is clearly unappropriate
for wind power data. The solid line is the fitted Gaussian distribution
by maximizing the likelihood.} \label{fig:epdf_log_bar_bw}
\end{figure}

\begin{figure}

\includegraphics{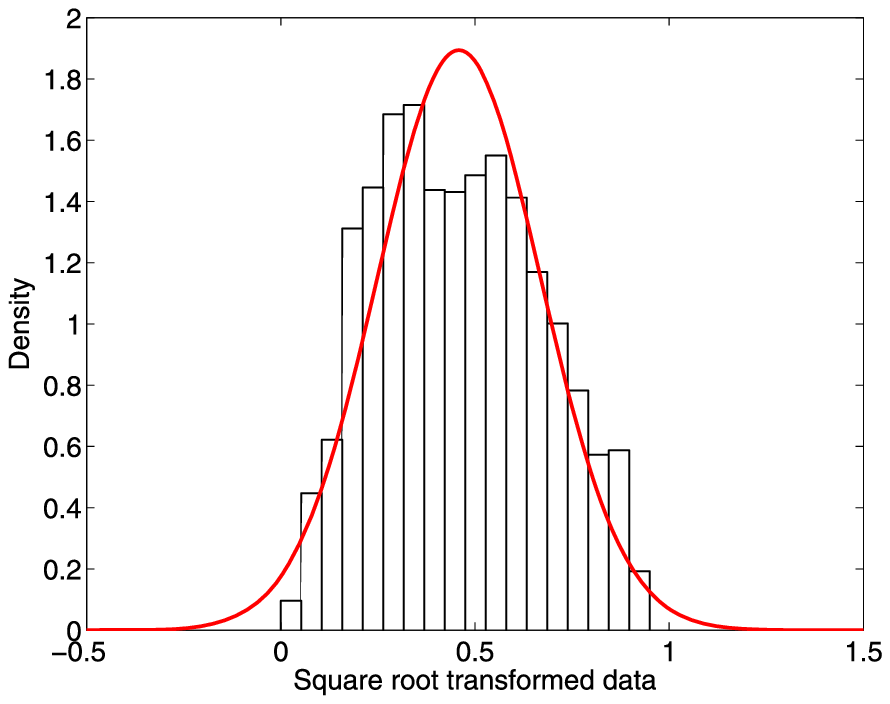}

\caption{Density of the wind power data after applying the square root
transformation, which remains non-Gaussian. The square root
transformation is a common transformation to convert wind speed data
into an approximate Gaussian distribution, but is clearly inappropriate
for wind power data. The solid line is the fitted Gaussian distribution
by maximizing the likelihood.}\label{fig:epdf_sqrt_bar_bw}
\end{figure}

\begin{figure}

\includegraphics{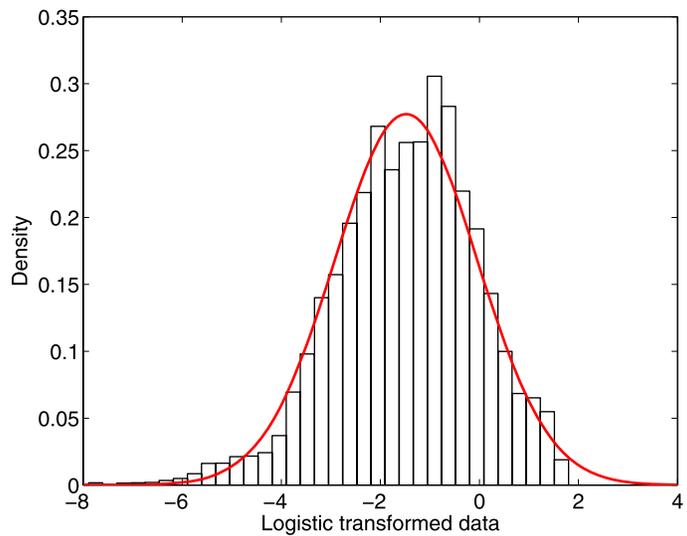}

\caption{Density of the wind power data after applying the logistic
transformation, which can be well approximated by a Gaussian
distribution. The solid line is the fitted Gaussian distribution by
maximizing the likelihood.} \label{fig:epdf_logit_bar_bw}
\end{figure}

Aggregated wind power time series, although smoother than those from individual wind farms, are non-Gaussian. In particular, they are nonnegative.
Figure~\ref{fig:epdf_bar_bw} shows the unconditional density of aggregated
wind power. This distribution has a sharper peak than the normal
distribution and is also significantly right-skewed. Common
transformations for normalizing wind speed data include the logarithmic
transformation and the square root transformation
[\citet{Taylor2006}]. However, those transformations are shown to
be unsatisfactory for our particular wind power data as demonstrated in Figures
\ref{fig:epdf_log_bar_bw} and \ref{fig:epdf_sqrt_bar_bw}. Nevertheless,
we could transform the wind power data $y_t$ by a logistic
transformation. This can be traced back to the work of
\citet{Johnson1949}, and recently \citet{Bremnes2006} applies
this transformation to model wind power. The logistic transformation is
given by
\begin{equation}\label{eq:logit}
  z_t = \log  \biggl( \frac{y_t}{1 - y_t}  \biggr), \qquad   0 < y_t < 1,
\end{equation}
and the transformed data $z_t$ gives a distribution which can be well
approximated by a Gaussian distribution as shown in Figure
\ref{fig:epdf_logit_bar_bw}. In contrast with individual wind power
data, we do not encounter any values of zero or one and so
(\ref{eq:logit}) is well defined. In Section \ref{sec:model} we apply
this transformation and build a Gaussian model to generate multi-step
density forecasts for wind power.

\section{Approaches for density forecasting}\label{sec:Approach}

Since our aim of this paper is to generate multi-step ahead density
forecasts without relying on Monte Carlo simulations, it is important
that our approach can be iterated easily. For this reason, in
both of the following approaches, we consider a Gaussian model at
certain stages so that we can iterate the forecasts in a tractable
manner.

\subsection{Gaussian model for transformed data}\label{sec:model}

In the first approach, we consider the transformation of wind power
data into an approximately Gaussian distribution so that we could
describe the transformed data by a simple Gaussian model, in
particular, the conventional ARIMA--GARCH model with Gaussian
innovations. As discussed in Section \ref{sec:data}, we transform the
wind power data by the logistic function in (\ref{eq:logit}). This
transformation maps the support from $(0,1)$ to the entire real axis,
and Figure \ref{fig:epdf_logit_bar_bw} shows that this results in an
approximately Gaussian distribution.

\begin{figure}

\includegraphics{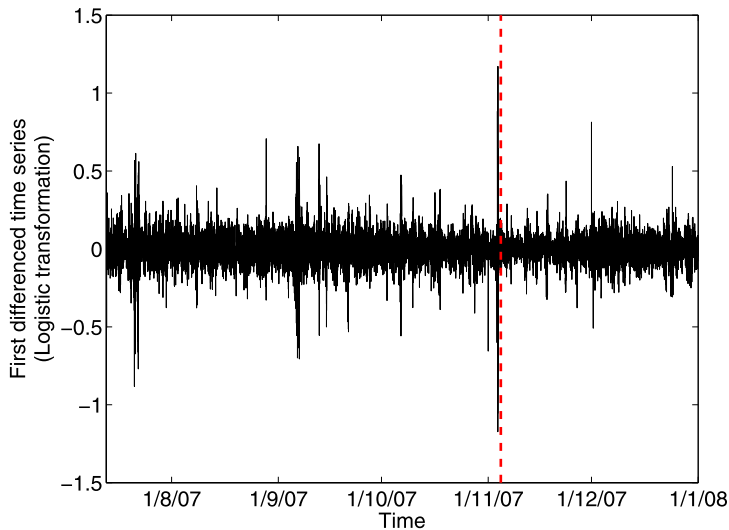}

\caption{First differences of the logistic transformed wind power. The
variance is not changing as fast as before, and the amount of
volatility clusterings is reduced. However, the time series is still
nonstationary. The data are dissected by the dashed line into a
training set and a testing set.} \label{fig:NormWP_logit_Diff_bw}
\end{figure}

\begin{figure}

\includegraphics{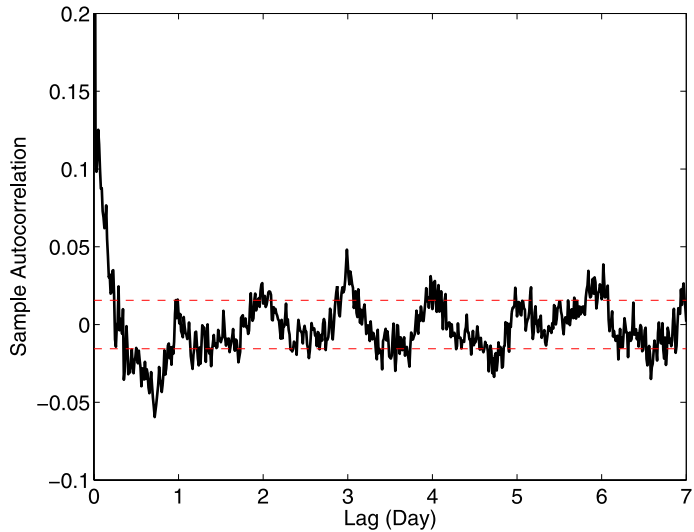}

\caption{Sample ACF of the first differences of logistic transformed
wind power up to a lag of 7 days. The
dashed lines are the confidence bounds at 2
standard deviations, assuming that the data
follow a Gaussian white noise process. The autocorrelations are
slightly smaller than that for the original data, which is shown in
Figure \protect\ref{fig:NormWPDiff_ACF_bw}.} \label{fig:NormWPLogitDiff_ACF_bw}
\end{figure}

As wind power data are nonstationary, so are the transformed data and
we consider the first differences $w_t = z_t - z_{t-1}$. When compared
with the original first differences $y_t - y_{t-1}$ in Figure
\ref{fig:NormWPDiff_bw}, the logistic transformed values $z_t$ have
fewer volatility clusterings and a smaller autocorrelation. This is
shown in Figure \ref{fig:NormWP_logit_Diff_bw} and Figure
\ref{fig:NormWPLogitDiff_ACF_bw}, respectively. Thus, we model $z_t$ by
an $\operatorname{ARIMA}(p,1,q$)--$\operatorname{GARCH}(r,s)$ model\footnote{We have also considered
modeling $z_t$ by $\operatorname{ARMA}(p,q$)--$\operatorname{GARCH}(r,s)$ models, but they are not
selected based on the BIC values.}
\begin{eqnarray}
w_t &=& \mu + \sum_{i=1}^{p} \phi_i w_{t-i} + \sum_{j=1}^{q} \theta_j \varepsilon_{t-j} + \varepsilon_t,\qquad
\varepsilon_t|\mathcal{F}_{t-1} \stackrel{\mathrm{i.i.d.}}{\sim} N(0,\sigma^2_{\varepsilon;t}),
\nonumber\\[-8pt]\\[-8pt]
\sigma^2_{\varepsilon;t} &=& \omega + \sum_{i=1}^{r} \alpha_i \varepsilon_{t-i}^2 + \sum_{j=1}^{s} \beta_j \sigma^2_{\varepsilon;t-j},  \nonumber
\end{eqnarray}
where $w_t = z_t - z_{t-1}$, $\mu, \phi_i, \theta_j, \omega, \alpha_i,
\beta_j$ are constant coefficients satisfying the usual conditions [Tsay (\citeyear{Tsay2005})] and
$\mathcal{F}_t$ consists of all the past values of $z$ up to time $t$.
We also consider an $\operatorname{ARIMA}(p,1,q$) model for $z_t$ with
constant conditional \mbox{variance}
$\operatorname{Var}[\varepsilon_t|\mathcal{F}_{t-1}] =
\sigma^2_{\varepsilon;t} = \sigma^2_\varepsilon$, so as to compare with
the $\operatorname{ARIMA}(p,1,q$)--$\operatorname{GARCH}(r,s$) model. We
select the models by minimizing the Bayesian Information Criteria
(BIC). Parameters are estimated by maximizing the Gaussian likelihood.

The optimal $h$-step ahead forecasts $\hat{z}_{t+h|t}$ and
$\hat{\sigma}^2_{\varepsilon;t+h|t}$ can be easily obtained, and the
corresponding $h$-step ahead density forecast of $Z_{t+h}$ is given by
the Gaussian distribution, that is, $f_{Z_{t+h|t}} \sim
N(\hat{z}_{t+h|t}, \hat{\sigma}^2_{t+h|t})$ so that\break
$\hat{\sigma}^2_{t+h|t} = \operatorname{Var}[z_{t+h}|\mathcal{F}_t]$
can be obtained from $\{ \hat{\sigma}^2_{\varepsilon;t+j|t} \}_{j=1}^h$
in a standard way,\vspace*{1pt} for example, by
expressing the model in a moving average
(MA) representation [Tsay (\citeyear{Tsay2005})]. To restore the density of the normalized aggregated
wind power $Y_{t+h}$, we compute the Jacobian of the transformation in
(\ref{eq:logit}) where $|J| =  | dz/dy  | = 1/[y(1-y)]$. The density of
$Y_{t+h}$ is then given by $f_{Y_{t+h|t}}(y_{t+h}) = |J| f_{Z_{t+h|t}}
(z_{t+h})$, that is,
\begin{eqnarray}\label{eq:LogitDensity}
    f_{Y_{t+h|t}}(y_{t+h}) &=& \frac{1}{y_{t+h} (1-y_{t+h})} \frac{1}{\sqrt{2 \pi
    \hat{\sigma}^2_{t+h|t}}}\nonumber\\[-8pt]\\[-8pt]
&&{}\times\exp   \biggl[ \biggl( -  \biggl( \log  \biggl( \frac{y_{t+h}}{1-y_{t+h}}  \biggr) - \hat{z}_{t+h|t}  \biggr)^2 \biggr)
                            \big/(2 \hat{\sigma}^2_{t+h|t})  \biggr].\nonumber
\end{eqnarray}
Note that (\ref{eq:LogitDensity}) is the $h$-step ahead conditional
density of $Y_{t+h}$ given the conditional point forecast of
$\hat{z}_{t+h|t}$ at time $t$.

\subsection{Exponential smoothing and truncated normal distribution}\label{sec:ESTrunNorm}

The second approach deals with the original wind power data $y_t$
directly. However, since the data are non-Gaussian, there is a problem
with the iteration of multi-step ahead density forecasts. We handle
this by expressing the $h$-step ahead conditional density as a function
of its first two moments. For instance, the one-step ahead density is
written as $f_{t+1|t}(y ; \hat{\mu}_{t+1|t}, \hat{\sigma}^2_{t+1|t})$,
where $\hat{\mu}_{t+1|t} = \mathrm{E}[y_{t+1} | \mathcal{F}_t]$ is the
conditional mean and $\hat{\sigma}^2_{t+1|t} =
\operatorname{Var}[y_{t+1} | \mathcal{F}_t] =
\operatorname{Var}[\varepsilon_{t+1} | \mathcal{F}_t] =
\hat{\sigma}^2_{\varepsilon;t+1|t}$ is the conditional variance.\footnote{In this paper, $\hat{\sigma}^2_{t+h|t}$ denotes the
conditional variance of the data $y_{t+h}$, while
$\hat{\sigma}^2_{\varepsilon;t+h|t}$ denotes the conditional variance
of the innovation $\varepsilon_{t+h}$, so that in general
$\hat{\sigma}^2_{t+h|t}$ is a function of
$\hat{\sigma}^2_{\varepsilon;t+j|t}$ with $j = 1,\ldots,h$.} At this
moment, we do not attempt to figure out the exact form of the density
function $f_{t+1|t}$. Given any $f_{t+1|t}$ and a model $M$ for the
dynamics, we can always evolve the density function so that
\begin{eqnarray}\label{eq:M}
  f_{t+1|t}(y ; \hat{\mu}_{t+1|t}, \hat{\sigma}^2_{t+1|t})  \stackrel{M}{\longrightarrow}
            f_{t+h|t}(y ; \hat{\mu}_{t+h|t}, \hat{\sigma}^2_{t+h|t}),\nonumber\\[1pt]\\[-21pt]
\eqntext{\hat{\mu}_{t+h|t} = p^{(h)}_M(\hat{\mu}_{t+1|t}, \ldots, \hat{\mu}_{t+h-1|t}; y_1, \ldots, y_t),}  \\
\eqntext{\hat{\sigma}^2_{t+h|t} = q^{(h)}_M(\hat{\sigma}^2_{\varepsilon;t+1|t}, \ldots, \hat{\sigma}^2_{\varepsilon;t+h|t}),\hspace*{51pt}}
\end{eqnarray}
where $\stackrel{M}{\longrightarrow}$ denotes the process of evolving
the dynamics and generating $h$-step ahead density forecasts under the
unknown model $M$, which in practice may require the use of Monte Carlo
simulations. Here $p^{(h)}_M$ and $q^{(h)}_M$ stand for functions that
give the conditional mean and the conditional variance of $y_t$, with
parameters that depend on the model $M$ and the forecast horizon $h$.

It is difficult to obtain any closed form for $f_{t+h|t}$ if the
distribution of innovations $\varepsilon_t$ is non-Gaussian. Thus, we
propose to use a two-step approach to approximate $f_{t+h|t}$. In the
first step, we attempt to model the dynamics of the conditional mean
$\hat{\ell}_{t+h|t}$ and the conditional variance $\hat{s}^2_{t+h|t}$
of the data using a Gaussian model~$G$. This is expressed as
\begin{eqnarray}\label{eq:M2}
\mbox{\textit{Step} 1:}\qquad \hat{\ell}_{t+h|t} &=& p^{(h)}_G(\hat{\ell}_{t+1|t}, \ldots, \hat{\ell}_{t+h-1|t}; y_1, \ldots,
y_t),
\nonumber\\[-8pt]\\[-8pt]
             \hat{s}^2_{t+h|t} &=& q^{(h)}_G(\hat{s}^2_{\varepsilon;t+1|t}, \ldots, \hat{s}^2_{\varepsilon;t+h|t}),  \nonumber
\end{eqnarray}
where $p^{(h)}_G$ and $q^{(h)}_G$ stand for functions that give the
conditional mean and the conditional variance of $y_{t+h}$, with
parameters that depend on the Gaussian model~$G$ and horizon $h$. In
model~$G$, the innovations are additive and are assumed to be i.i.d.
Gaussian distributed. For example, $G$ can be the conventional
ARIMA--GARCH model with Gaussian innovations. This may be violated in
reality, so $\hat{\ell}_{t+h|t}$ and $\hat{s}^2_{t+h|t}$ obtained from
model $G$ may not be the true conditional mean $\hat{\mu}_{t+h|t}$ and
conditional variance $\hat{\sigma}^2_{t+h|t}$ respectively. They only
serve as proxies to the true values.

Although model $G$ may not describe real situations, we rely on a
second step for remedial adjustments such that the final density
forecast is an approximation to reality. In the second step, we assume
that the $h$-step ahead density $f_{t+h|t}$ can be approximated by a
parametric function $D$, which is characterized by a location parameter
and a scale parameter. In particular, the location parameter and the
scale parameter are obtained from the conditional mean
$\hat{\ell}_{t+h|t}$ and the conditional variance $\hat{s}^2_{t+h|t}$
respectively, which are estimated from the Gaussian model $G$. Thus, we
simply take
\begin{equation}\label{eq:M2_2}
 \mbox{\textit{Step} 2:}\qquad f_{t+h|t}(y ; \hat{\mu}_{t+h|t}, \hat{\sigma}^2_{t+h|t}) \approx D (y ; \hat{\ell}_{t+h|t}, \hat{s}^2_{t+h|t})
\end{equation}
as the $h$-step ahead density forecast where $D$ is a function
depending on two parameters only. As a result, the two-step approach
may be able to give a good estimation of $f_{t+h|t}$ if (\ref{eq:M2_2})
is a close approximation. In (\ref{eq:M2_2}) the correct conditional
mean $\hat{\mu}_{t+h|t}$ and conditional variance
$\hat{\sigma}^2_{t+h|t}$ are generated by $p^{(h)}_M(\cdot)$ and
$q^{(h)}_M(\cdot)$ under the true model $M$, while the corresponding
proxy values $\hat{\ell}_{t+h|t}$ and $\hat{s}^2_{t+h|t}$ are generated
by $p^{(h)}_G(\cdot)$ and $q^{(h)}_G(\cdot)$ under a Gaussian model
$G$. Empirical studies will be needed to determine the appropriate
Gaussian model $G$ as well as the best choice $D$ in order to
approximate the final density $f_{t+h|t}$.

For our normalized aggregated wind power $y_t$, choosing $D$ as the
truncated normal distribution bounded within $[0,1]$ gives a good
approximation of $f_{t+h|t}$. Truncated normal distributions have been
applied successfully in modeling bounded, nonnegative data
[\citet{Sanso1999}, \citet{Gneiting2006}]. We consider $D$ to be
parameterized by the location parameter $\hat{\ell}_{t+h|t}$ and the
scale parameter $\hat{s}^2_{t+h|t}$, where $N(\hat{\ell}_{t+h|t},
\hat{s}^2_{t+h|t})$ is the corresponding normal distribution without
truncation. Note that $\hat{\ell}_{t+h|t}$ and $\hat{s}^2_{t+h|t}$ will
be the true conditional mean and conditional variance if the data are
indeed Gaussian. The density function $f_{t+h|t} $ is then given by
(\ref{eq:M2_2}) so that
\begin{eqnarray}\label{eq:TrunNormDensity}
    f_{t+h|t}(y; \hat{\mu}_{t+h|t}, \hat{\sigma}^2_{t+h|t}) &=&
    \frac{1}{\hat{s}_{t+h|t}}
    \biggl( \varphi  \biggl( \frac{y-\hat{\ell}_{t+h|t}}{\hat{s}_{t+h|t}}  \biggr) \biggr)\nonumber\\[-8pt]\\[-8pt]
    &&{}\Big/
    \biggl(\Phi  \biggl( \frac{1-\hat{\ell}_{t+h|t}}{\hat{s}_{t+h|t}}  \biggr) - \Phi  \biggl( \frac{-\hat{\ell}_{t+h|t}}{\hat{s}_{t+h|t}}
    \biggr)\biggr)\nonumber
\end{eqnarray}
for $y\in(0,1)$, where $\varphi$ and $\Phi$ are the standard normal density and distribution function respectively.

Instead of directly estimating $\hat{\ell}_{t+h|t}$ and
$\hat{s}^2_{t+h|t}$ using the ARIMA--GARCH models, we find that a
better way is to smooth the two parameters simultaneously by
exponential smoothing methods. Exponential smoothing methods have been
widely and successfully adopted in areas such as inventory forecasting
[\citet{Brown1961}], electricity forecasting
[\citet{Taylor2003}] and volatility forecasting
[\citet{Taylor2004}]. A comprehensive review of exponential
smoothing is given by \citet{Gardner2006}.
\citet{Hyndman2008} provide a state space framework for
exponential smoothing, which further strengthens its value as a
statistical model instead of an ad hoc forecasting procedure.
\citet{Ledolter1978} show that exponential smoothing methods
produce optimal point forecasts if and only if the underlying data
generating process is within a subclass of
$\operatorname{ARIMA}(p,d,q$) processes. We extend this property and
demonstrate that simultaneous exponential smoothing on the mean and
variance can produce optimal point forecasts if the data follow a
corresponding $\operatorname{ARIMA}(p,d,q$)--$\operatorname{GARCH}(r,s$)
process. This enables us to generate multi-step ahead forecasts for the
parameters $\hat{\ell}_{t+h|t}$ and $\hat{s}^2_{t+h|t}$ by iterating
the underlying ARIMA--GARCH model of exponential smoothing.

\subsubsection{Smoothing the location parameter only}\label{sec:ETSmean}

For the simplest case, let us assume that the conditional variance
of wind power is constant. This means that we only need to smooth the
conditional mean $\ell_t$, while the conditional variance $s^2_t$
will be estimated directly from the data via estimating the variance of
innovations $\hat{s}_{\varepsilon}^2$. From now on, we refer to the
conditional mean as the location parameter and the conditional variance
as the scale parameter so as to remind us that they correspond to the
truncated normal distribution. Again, the $h$-step ahead scale
parameter $\hat{s}_{t+h|t}^2$ is obtained as a function of
$\hat{s}_{\varepsilon}^2$.

By simple exponential smoothing, the smoothed series of the location
parameter $\ell_t$ is given by $S_t$, which is updated according to
\begin{equation}\label{eq:ESupdate}
    S_t = \alpha y_t + (1-\alpha)S_{t-1},
\end{equation}
where $y_t$ is the observed wind power at time $t$ and $0<\alpha<1$ is
a smoothing parameter. We initialize the series by setting $S_1 = y_1$,
and the one-step ahead forecast is $\hat{\ell}_{t+1|t} = S_t$.
Iterating (\ref{eq:ESupdate}) gives $\hat{\ell}_{t+h|t} = S_t$.
However, the forecast errors $y_t - \hat{\ell}_{t|t-1}$ are highly
correlated, with a significant lag one sample autocorrelation of
$0.2723$. A simple way to improve the forecast is to add a parameter
$\phi_s$ to account for autocorrelations in the forecast equation
[\citet{Taylor2003}]. We call this the simple exponential
smoothing with error correction. The updating equation is still given
by (\ref{eq:ESupdate}), but the forecast equation is modified as
\begin{equation}\label{eq:ESNNECforecast}
    \hat{\ell}_{t+1|t} = S_t + \phi_s (y_t - S_{t-1}),
\end{equation}
where $|\phi_s|<1$. Note that it is now possible to obtain negative
values for $\hat{\ell}_{t+1|t}$ in (\ref{eq:ESNNECforecast}) and in
such cases $\hat{\ell}_{t+1|t}$ is obviously not the true conditional
mean. Nevertheless, this is not a problem here since
$\hat{\ell}_{t+1|t}$ essentially serves as the location parameter of
the truncated normal distribution, which can be negative. Following the
taxonomy introduced by \citet{Hyndman2008}, we denote
(\ref{eq:ESupdate}) and (\ref{eq:ESNNECforecast}) as the
$\operatorname{ETS}(A,N,N|\mathit{EC})$ method, where ETS stands for both an abbreviation for
exponential smoothing as well as an acronym for error, trend and
seasonality respectively. The $A$ inside the bracket stands for
additive errors in the model, the first $N$ stands for no trend, the
second $N$ stands for no seasonality and $\mathit{EC}$ stands for error
correction.

By directly iterating (\ref{eq:ESupdate}) and (\ref{eq:ESNNECforecast})
and expressing $\hat{y}_{t+h|t} = \hat{\ell}_{t+h|t}$, we have
\begin{equation}\label{ESNNECforecast_h}
    \hat{\ell}_{t+h|t}
    = S_t + \frac{\alpha \phi_s (1 - \phi_s^{h-1})}{1-\phi_s}(y_t - S_{t-1}) + \phi_s^h (y_t - S_{t-1})
\end{equation}
for $h>1$. To generate $h$-step ahead forecasts of $\hat{s}_{t+h|t}^2$,
it is important that we identify an underlying model corresponding to
our updating and forecast equations (\ref{eq:ESupdate}) and
(\ref{eq:ESNNECforecast}). It can be easily checked that the
$\operatorname{ETS}(A,N,N|\mathit{EC})$ method is optimal for the
$\operatorname{ARIMA}(1,1,1)$ model, in the sense that the forecasts in
(\ref{eq:ESNNECforecast}) are the minimum mean square error (MMSE)
forecasts. Expressed in the form of an $\operatorname{ARIMA}(1,1,1)$ model with Gaussian
innovations, the $\operatorname{ETS}(A,N,N|\mathit{EC})$ method can be
written as
\begin{equation}\label{eq:ARIMA111_ETS}
    w_t = \phi_s w_{t-1} + \varepsilon_t + (\alpha - 1)\varepsilon_{t-1}, \qquad\varepsilon_t \stackrel{\mathrm{i.i.d.}}{\sim} N(0,s_{\epsilon}^2),
\end{equation}
where $w_t = y_t - y_{t-1}$, $\varepsilon_t$ is the Gaussian innovation
with mean zero and constant variance $s_{\epsilon}^2$, and $\alpha,
\phi_s$ are the smoothing parameters in (\ref{eq:ESupdate}) and
(\ref{eq:ESNNECforecast}). It can also be easily verified that
(\ref{ESNNECforecast_h}) is identical to the $h$-step ahead forecasts
obtained from the $\operatorname{ARIMA}(1,1,1)$ model in
(\ref{eq:ARIMA111_ETS}). It then follows from the
$\operatorname{ARIMA}(1,1,1)$ model that the $h$-step ahead forecast
variance is given by
\begin{equation}\label{ESNNECforecast_sig_h}
    \hat{s}^2_{t+h|t} = \hat{s}_{\epsilon}^2 \sum_{j=1}^h \Omega_{h-j}^2,
\end{equation}
where $\Omega_{0} = 1, \Omega_{h} = \phi_s^h + \alpha
(1-\phi_s^h)/(1-\phi_s)$ for $h\geq1$, and $\hat{s}^2_{\varepsilon}$ is
the estimated constant variance of the innovations. Note that in this
case, (\ref{ESNNECforecast_sig_h}) is the explicit form of
$\hat{s}^2_{t+h|t} = q^{(h)}_G(\hat{s}_{\epsilon}^2)$ in (\ref{eq:M2}).

Since maximum likelihood estimators are well known to have nice
asymptotic properties, we estimate the three parameters $\alpha,
\phi_s$ and $\hat{s}^2_{\varepsilon}$ by maximizing the likelihood of
the truncated normal distribution $f_{t+1|t}(y_{t+1};
\hat{\ell}_{t+1|t}, \hat{s}^2_{t+1|t})$. One may also consider
minimizing the mean continuous ranked probability scores (CRPS) of the
density forecasts [\citeauthor{Gneiting2005} (\citeyear{Gneiting2005}, \citeyear{Gneiting2006})], but this
requires a much larger amount of computation. Although it may slightly
improve the density forecasts, minimizing the CRPS is not appealing
here since we aim at generating multi-step forecasts in a
computationally efficient way. After obtaining the parameters, from
(\ref{ESNNECforecast_h}) and (\ref{ESNNECforecast_sig_h}) we can
generate the $h$-step ahead density forecasts using
(\ref{eq:TrunNormDensity}).

\subsubsection{Smoothing both the location and scale parameters simultaneously}

Next, we consider heteroscedasticity for the conditional variances of
wind power. In this case, apart from smoothing the location parameter
$\ell_t$, we also simultaneously smooth the scale parameter $s_t^2$. In
fact, we smooth the variance of innovations $s_{\varepsilon;t}^2$ and
obtain the scale parameter $s_t^2$ as a function of
$s_{\varepsilon;t}^2$ as in (\ref{eq:M2}).

Equipped with the one-step ahead forecast of the location parameter
$\hat{\ell}_{t|t-1}$, we may calculate the squared difference between
$\hat{\ell}_{t|t-1}$ and the observed wind power $y_t$, that is, $(y_t
- \hat{\ell}_{t|t-1})^2$, as the estimated variance
$s^2_{\varepsilon;t}$ at time $t$. Applying simple exponential
smoothing, the smoothed series of $s_{\varepsilon;t}^2$ is given by
$V_t$, which is updated according to
\begin{equation}\label{eq:ESVarupdate}
    V_t = \gamma (y_t - \hat{\ell}_{t|t-1})^2 + (1-\gamma) V_{t-1},
\end{equation}
where $y_t$ is the observed wind power at time $t$,
$\hat{\ell}_{t|t-1}$ is obtained by (\ref{eq:ESNNECforecast}) and
$0<\gamma<1$ is a smoothing parameter. We initialize the series by
setting $V_1$ to be the variance of the data in the training set. In
fact, the forecasts are not sensitive to the choice of initial values
due to the size of the data set. The one-step ahead forecast is given
by $\hat{s}^2_{\varepsilon;t+1|t} = V_t$. Again, the forecast errors
are highly correlated and it is better to include an additional
parameter $\phi_v$ in the forecast equation to account for
autocorrelations. The modified forecast equation is then given by
\begin{equation}\label{eq:ESVarECforecast}
    \hat{s}^2_{\varepsilon;t+1|t} = V_t + \phi_v  [ (y_t - \hat{\ell}_{t|t-1})^2 - V_{t-1}
    ],
\end{equation}
where $|\phi_v|<1$. Unfortunately, a major drawback of introducing this
extra term in the forecast equation is that negative values of
$\hat{s}^2_{\varepsilon;t+1|t}$ may occur. Although this does not
happen in our data,\vspace*{1pt} we modify our approach and consider smoothing the
logarithmic transformed scale parameter $\log s^2_{\varepsilon;t}$ such
that negative values are allowed since we aim at developing a general
methodology that applies to different data sets. The smoothed series
for $\log s^2_{\varepsilon;t}$ is then given by $\log V_t$. Denoting\vspace*{-3pt}
$\varepsilon_t = y_t - \hat{\ell}_{t|t-1}$ and $e_t = \varepsilon_t /
\sqrt{V_t}$, the estimated logarithmic variance at time $t$ is now chosen to be
$g(e_{t})$ instead of $\log\varepsilon_t^2$ so that
\begin{equation}\label{eq:g}
    g(e_t) = \theta  ( |e_t| - \mathrm{E}[|e_t|]  ),
\end{equation}
where $\theta$ is a constant parameter. This ensures that
$g( e_t)$  is positive for large values of $e_t$ and
negative if $e_t$ is small. The updating equation and the forecast equation are now written respectively as
\begin{eqnarray}\label{eq:ESlogVarEqn}
     \log V_t &=& \gamma g(e_{t}) + (1-\gamma) \log V_{t-1},
     \nonumber\\[-8pt]\\[-8pt]
    \log \hat{s}^2_{\varepsilon;t+1|t} &=& \log V_t + \phi_v  [ g(e_{t}) - \log V_{t-1}  ],\nonumber
\end{eqnarray}
which are analogous to (\ref{eq:ESVarupdate}) and
(\ref{eq:ESVarECforecast}), except that a logarithmic\vspace*{1pt} transformation is
taken and $(y_t - \hat{\ell}_{t|t-1})^2$ is replaced by $g(e_{t})$.
We initialize the series by setting $\log V_1 = 0$. In fact, the
smoothing procedure is insensitive to the initial value due to the size
of the data set.

Now, the $h$-step ahead forecasts of $\hat{\ell}_{t+h|t}$ are still obtained from (\ref{ESNNECforecast_h}), but to generate $h$-step ahead forecasts of $\hat{s}_{t+h|t}^2$ we need to identify an underlying model for this smoothing method. We summarize our exponential smoothing method for both $\ell_t$ and $s_t^2$ by combining (\ref{eq:ESupdate}), (\ref{eq:ESNNECforecast}) and (\ref{eq:ESlogVarEqn}):
\begin{eqnarray}\label{eq:ESNNEC_ECforecast}
S_t &=& \alpha y_t + (1-\alpha)S_{t-1} ,  \nonumber \\
\hat{\ell}_{t+1|t} &=& S_t + \phi_s (y_t - S_{t-1}),\nonumber\\[-8pt]\\[-8pt]
\log V_t &=& \gamma g(e_{t}) + (1-\gamma) \log V_{t-1} ,   \nonumber \\
\log \hat{s}^2_{\varepsilon;t+1|t} &=& \log V_t + \phi_v  [ g(e_{t}) - \log V_{t-1}  ],\nonumber
\end{eqnarray}
where $g(e_t)$ is given in (\ref{eq:g}) and $e_t$ as defined
previously. There are four smoothing parameters $\alpha, \gamma,
\phi_s, \phi_v$ and a parameter $\theta$ for the estimated logarithmic variance
$g(e_t)$. We adopt the taxonomy similar to that for exponential
smoothing for the location parameter as described in Section
\ref{sec:ETSmean}, and denote (\ref{eq:ESNNEC_ECforecast}) as the
$\operatorname{ETS}(A,N,N|\mathit{EC})$--($A,N,N|\mathit{EC}$) method where
the second bracket of $(A,N,N|\mathit{EC})$ indicates the exponential
smoothing method applied for smoothing the variance. We aim at
identifying (\ref{eq:ESNNEC_ECforecast}) with an ARIMA--GARCH model.
Using (\ref{eq:ARIMA111_ETS}) as the $\operatorname{ARIMA}(1,1,1)$
model for $y_t$ and  writing $\varepsilon_t = y_t -
\hat{\ell}_{t|t-1}$, the last equation in (\ref{eq:ESNNEC_ECforecast})
can be written as
\begin{eqnarray}
\log \hat{s}^2_{\varepsilon;t+1|t}  &=& \log V_t + \phi_v [ g(e_{t}) - \log V_{t-1} ] \nonumber \\
&=& \gamma g(e_{t}) + (1-\gamma) \log V_{t-1} + \phi_v [ g(e_{t}) - \log V_{t-1} ]\nonumber\\
&=& (\gamma + \phi_v) g(e_{t}) - \phi_v \log V_{t-1} \\
&&{} + (1-\gamma) \{ \log s^2_{\varepsilon;t} - \phi_v [ g(e_{t-1}) - \log V_{t-2} ] \} \nonumber\\
&=& (\gamma + \phi_v) g(e_{t}) - \phi_v g(e_{t-1}) + (1-\gamma) \log s^2_{\varepsilon;t},\nonumber
\end{eqnarray}
where we have used the updating equation in (\ref{eq:ESlogVarEqn}).
This is the exponential GARCH, that is, $\operatorname{EGARCH}(2,1)$ model for the
conditional variance of innovations $\varepsilon_t$
[\citet{Nelson1991}]. Unlike the conventional EGARCH models for
asset prices, $g(e_t)$ is symmetric since there is no reason to expect
volatility to increase when wind power generation drops. In summary,
the exponential smoothing method in (\ref{eq:ESNNEC_ECforecast}) is
optimal for the $\operatorname{ARIMA}(1,1,1)$--$\operatorname{EGARCH}(2,1)$ model, which
can be written as
\begin{eqnarray}\label{eq:ARIMA111GARCH21}
  w_t &=& \phi_s w_{t-1} + \varepsilon_t + (\alpha -
    1)\varepsilon_{t-1},\qquad
                \varepsilon_t | \mathcal{F}_{t-1} \stackrel{\mathrm{i.i.d.}}{\sim}
                N(0,s^2_{\varepsilon;t}),\nonumber\\[-8pt]\\[-8pt]
    \log s_{\varepsilon;t}^2 &=& (1-\gamma) \log s_{\varepsilon;t-1}^2 + (\gamma + \phi_v) g(e_{t-1}) - \phi_v g(e_{t-2}),   \nonumber
\end{eqnarray}
where $w_t = y_t - y_{t-1}$ and $g(e_t)$ is given in (\ref{eq:g}), and
we have assumed Gaussian innovations so that $\mathrm{E}[|e_t|] =
\sqrt{2/\pi}$. Similarly, we estimate the five parameters $\alpha,
\phi_s, \gamma, \phi_v$ and $\theta$ by maximizing the truncated normal
likelihood as mentioned in Section \ref{sec:ETSmean}. Now, equipped with the
$\operatorname{ARIMA}(1,1,1)$--$\operatorname{EGARCH}(2,1)$ model in
(\ref{eq:ARIMA111GARCH21}), the $h$-step ahead forecasts for the scale
parameter $\hat{s}^2_{\varepsilon;t+h|t}$ can be easily obtained
[\citet{Tsay2005}]. Consequently, the $h$-step ahead forecasts
$\hat{s}^2_{t+h|t}$ can be expressed as a function of
$\{\hat{s}^2_{\varepsilon;t+j|t} \}_{j=1}^{h}$, which is analogous to\vspace*{2pt}
(\ref{ESNNECforecast_sig_h}) except that the expression is much more
complicated and, in practice, one would simply iterate the forecasts.
The $h$-step ahead density forecasts can then be obtained using
(\ref{eq:TrunNormDensity}).

\section{Forecast evaluations}\label{sec:Evaluation}

\subsection{Benchmark models}

In this section we apply the approaches of density forecasts in Section
\ref{sec:Approach} to forecast normalized aggregated wind power in
Ireland. To evaluate the forecast performances of our approaches, we
compare the results with four simple benchmarks. The first two
benchmarks are the persistence (random walk) forecast and the constant
forecast, which are both obtained as truncated normal distributions in
(\ref{eq:TrunNormDensity}). For the persistence forecast, we estimate
the $h$-step ahead location parameter $\hat{\ell}_{t+h|t}$ and scale
parameter $\hat{s}_{t+h|t}^2$ using the latest observations, that is,
\begin{equation}\label{eq:persistence}
    \hat{\ell}_{t+h|t} = y_t,\qquad
                \hat{s}_{t+h|t}^2 = \frac{\sum_{j=1}^N (y_{t+1-j} - y_{t-j})^2}{N}
\end{equation}
for $t>N$. We find that taking $N=48$, that is, using data in the past 12 hours, gives an appropriate estimate for $\hat{s}_{t+h|t}^2$.

For the constant forecast, we estimate the constant location parameter
$\hat{\ell}_{t+h|t}$ and scale parameter $\hat{s}_{t+h|t}^2$ using data
in the whole training set. They are given by the sample mean and the
sample variance of the 11,008 observations in the training set, so that
\begin{equation}
    \hat{\ell}_{t+h|t} = \hat{\ell} = \frac{\sum_{j=1}^{11{,}008}
    y_{j}}{11{,}008},\qquad
                \hat{s}_{t+h|t}^2 = \hat{s}^2 = \frac{\sum_{j=1}^{11{,}008} (y_{j} - \hat{\ell})^2}{11{,}007}.
\end{equation}
We have also considered generating the persistence and constant
forecasts using the first approach as described in Section
\ref{sec:model}. However, our results show that the second approach
gives a better benchmark in terms of forecast performance.

\begin{figure}

\includegraphics{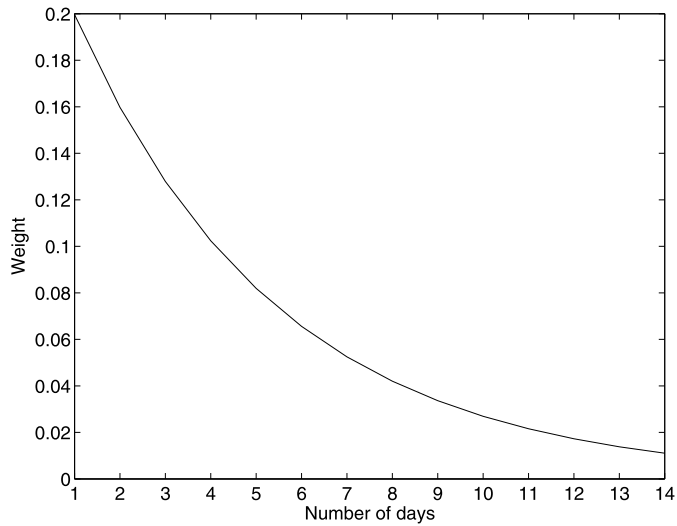}

\caption{The exponential decrease of the weights
$\lambda(1-\lambda)^{j-1}$ assigned to the empirical conditional
densities $g_{\mathrm{emp}}(\{ \Lambda_t^{j} \})$ fitted with $j$ days
of latest observations, where $\lambda = 0.1988$ is obtained by
maximizing the likelihood using data in the training set. The EWMA
empirical conditional density forecasts are obtained as the weighted
average of $g_{\mathrm{emp}}(\{ \Lambda_t^{j} \})$.}
\label{fig:EWMACondDen}
\end{figure}

On the other hand, the third and the fourth benchmarks are obtained by
estimating empirical densities from the data. The third benchmark is
the climatology forecast, in which an empirical unconditional density
is fitted using data in the whole training set. The density has been
shown in Figure \ref{fig:epdf_bar_bw} previously. The fourth benchmark
is the empirical conditional density forecast. To be in line with the
use of exponential smoothing to estimate the location and scale
parameters in Section \ref{sec:ESTrunNorm}, we consider an
exponentially weighted moving average (EWMA) of a set of empirical
conditional densities. Due to computational efficiency as well as
reliability of density estimations, at each time $t$ we essentially
consider the EWMA of 14 empirical conditional densities
$g_{\mathrm{emp}}(\{ \Lambda_t^{j} \})$, where each of them is fitted
using observations in the past $j$ days with $j = 1,2,\ldots,14$ and
$\{ \Lambda_t^{j} \} = \{ y_{t-96j+1}, y_{t-96j+2}, \ldots, y_{t} \}$
is the set of $(96 \times j)$ latest observations used to fit the
empirical density. Up to an appropriate normalization constant, the
$h$-step ahead EWMA empirical conditional density forecast is given by
\begin{equation}\label{eq:EWMACondDen}
    f_{t+h|t}(y) \propto \sum_{j=1}^{14} \lambda (1 - \lambda)^{j-1} g_{\mathrm{emp}}(\{ \Lambda_t^{j} \})
\end{equation}
so that for any fixed forecast origin $t$, the $h$-step ahead density
forecasts are identical for all $h>1$. The smoothing parameter in
(\ref{eq:EWMACondDen}) is estimated to be $\lambda = 0.1988$, which is
obtained by maximizing the log likelihood, that is, $\sum \log
f_{t+1|t}(\lambda; y_{t+1})$, using the data in the training set only.
It is possible to estimate a smoothing parameter for each forecast
horizon $h$. However, the improvements are not significant and, thus,
we simply keep using $\lambda = 0.1988$ for all horizons. Figure
\ref{fig:EWMACondDen} shows the exponential decrease of the weights
being assigned to different empirical densities $g_{\mathrm{emp}}(\{
\Lambda_t^{j} \})$.

In summary, we consider the following 4 benchmarks and 4 approaches of
generating multi-step density forecasts, and compare their forecast
performances from 15 minutes up to 24 hours ahead:
\begin{enumerate}
  \item Persistence forecast [TN]
  \item Constant forecast [TN]
  \item Climatology forecast [Empirical density]
  \item EWMA conditional density forecast [Empirical density]
  \item The $\operatorname{ARIMA}(2,1,3)$ model [LT]
  \item The $\operatorname{ARIMA}(4,1,3)$--$\operatorname{GARCH}(1,1)$ model [LT]
  \item The $\operatorname{ETS}(A,N,N|\mathit{EC})$ method [TN]
  \item The $\operatorname{ETS}(A,N,N|\mathit{EC})$--($A,N,N|\mathit{EC}$) method [TN],
\end{enumerate}
where [LT] stands for logistic transformation and [TN] stands
for truncated normal distribution, so as to remind us how the densities
are generated.

\subsection{Point forecasts}

First, let us evaluate the point forecasts generated by different
approaches. We consider the expected values of the density forecasts as
the optimal point forecasts. Given a forecast density, we can obtain
the expected value directly by numerical integration. In particular,
for forecast densities in the form of truncated normal distributions,
one may easily write down the expected value as
\begin{eqnarray}
    \hat{y}_{t+h|t} &=& \hat{\ell}_{t+h|t} - \hat{\ell}_{t+h|t} \biggl(
    \biggl(\varphi  \biggl(
    \frac{1-\hat{\ell}_{t+h|t}}{\hat{s}_{t+h|t}}  \biggr) - \varphi
    \biggl( \frac{-\hat{\ell}_{t+h|t}}{\hat{s}_{t+h|t}}  \biggr)
    \biggr)\nonumber\\[-8pt]\\[-8pt]
    &&\hphantom{ \hat{\ell}_{t+h|t} - \hat{\ell}_{t+h|t} \biggl(}
    {}\Big/ \biggl(\Phi  \biggl(
    \frac{1-\hat{\ell}_{t+h|t}}{\hat{s}_{t+h|t}}  \biggr) - \Phi
    \biggl( \frac{-\hat{\ell}_{t+h|t}}{\hat{s}_{t+h|t}}  \biggr)
    \biggr)  \biggr),\nonumber
\end{eqnarray}
where $\hat{\ell}_{t+h|t}$ and $\hat{s}^2_{t+h|t}$ are the location and
scale parameters of the truncated normal distribution in
(\ref{eq:TrunNormDensity}). Note that due to the truncation, the
distribution may not be symmetric and so the expected value is in
general different from the location parameter, that is,
$\hat{y}_{t+h|t} \neq \hat{\ell}_{t+h|t}$. In fact, referring to
(\ref{eq:M2}), $\hat{\ell}_{t+h|t} = p^{(h)}_G(\hat{\ell}_{t+1|t},
\ldots, \hat{\ell}_{t+h-1|t}; y_1, \ldots, y_t)$ is obtained according
to a Gaussian model $G$, which may not give the true conditional mean
$\hat{y}_{t+h|t}$ of the data, and may even be negative. Since the
final density $f_{t+h|t}$ is only obtained when an appropriate function
$D$ is chosen,\vspace*{1pt} we see that $D$ transforms the conditional mean from
$\hat{\ell}_{t+h|t}$ for Gaussian data to the optimal forecast
$\hat{y}_{t+h|t}$ for our data. This is analogous to calculating
optimal point forecasts when the loss function is asymmetric
[\citet{Christoffersen1997}, \citet{Patton2007}]. Since the normalized
aggregated wind power is bounded within $[0,1]$, the loss function is
always asymmetric unless the conditional mean is $\hat{\ell}_{t+h|t} =
0.5$. When the conditional mean is not the optimal forecast, an
additional term is added to compensate for the asymmetric loss.
\citet{Christoffersen1997} suggest an approximation to calculate
the optimal forecast for conditionally Gaussian data by assuming
$\hat{y}_{t+h|t} = G(\mu_{t+h|t}, \sigma^2_{t+h|t})$, where
$\mu_{t+h|t}, \sigma^2_{t+h|t}$ are the conditional mean and
conditional variance. Their method involves expanding $G$ into a Taylor
series.

To evaluate the performances of different forecasting approaches, we
calculate $h$-step ahead point forecasts for each of the 5504 values in
the testing set, where $1 \leq h \leq 96$, that is, from 15 minutes up
to 24 hours ahead. For each forecast horizon $h$, we calculate the mean
absolute error (MAE) and the root mean squared error (RMSE) of the
point forecasts, where the mean is taken over the 5504 $h$-step ahead
forecasts in the testing set.

\begin{figure}

\includegraphics{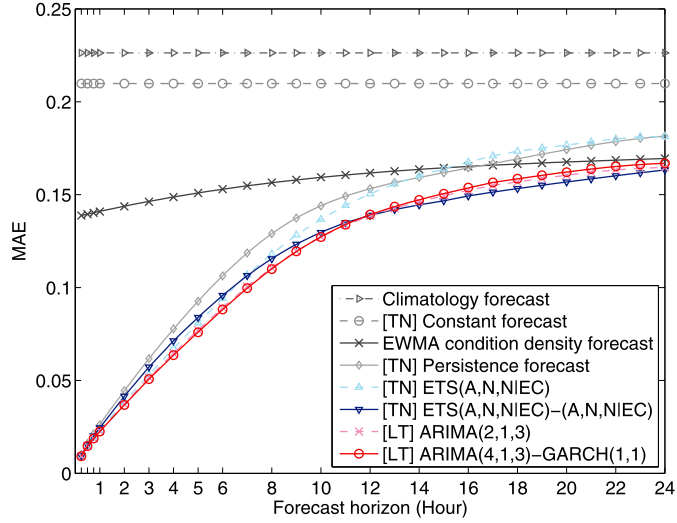}

\caption{Mean absolute error (MAE) of point forecasts generated by
different approaches for forecast horizons from 15 minutes to 24 hours
ahead. The ARIMA--GARCH models on logistic
transformed data perform best for short
horizons less than 12 hours whereas the
ETS$(A,N,N|EC)$--$(A,N,N|EC)$ method with
truncated normal distribution is best for
horizons greater than 12 hours.}
\label{fig:mae}
\end{figure}

\begin{figure}

\includegraphics{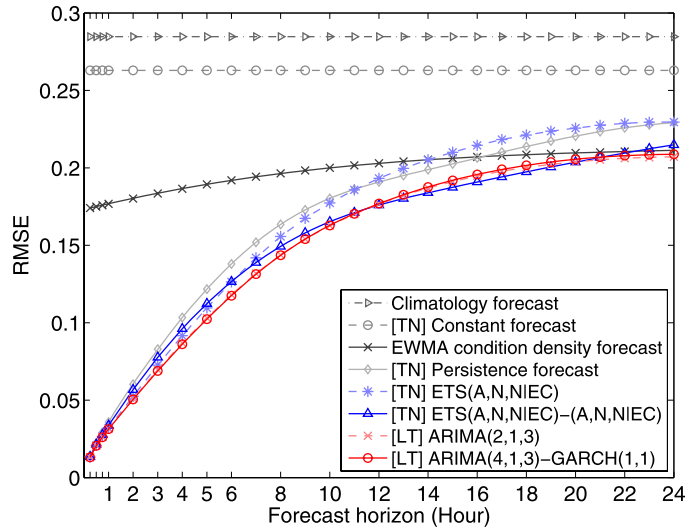}

\caption{Root mean squared error (RMSE) of point forecasts generated by
different approaches for forecast horizons from 15 minutes to 24 hours
ahead. Results are similar to those under MAE.} \label{fig:rmse}
\end{figure}

Figures \ref{fig:mae} and \ref{fig:rmse} show the results of point
forecasts under MAE and RMSE respectively. The rankings of different
approaches are similar under either MAE or RMSE, except for the
$\operatorname{ETS}(A,N,N|\mathit{EC})$--$(A,N,N|\mathit{EC})$ method which
performs relatively better under MAE than RMSE. It performs the best
under MAE for long forecast horizons beyond 14 hours. On the other
hand, the two ARIMA--GARCH models outperform all other approaches for
short forecast horizons within 12 hours, and are almost as good as the
$\operatorname{ETS}(A,N,N|\mathit{EC})$--$(A,N,N|\mathit{EC})$ method for
horizons beyond 12 hours.

Interestingly, the $\operatorname{ARIMA}(2,1,3)$ model is performing almost identically
to the $\operatorname{ARIMA}(4,1,3)$--$\operatorname{GARCH}(1,1)$ model. This phenomenon is in contrast
with that for the ETS methods, where smoothing both the location and
scale parameters do perform much better. It seems that including the
dynamics of the conditional variance in the modeling of the logistic
transformed wind power $z_t$ cannot improve the point forecasts under
MAE or RMSE. These may be explained by Figure \ref{fig:NormWPDiff_bw}
which shows a significantly changing variance in the original wind
power data $y_t$, and by Figure \ref{fig:NormWP_logit_Diff_bw} which
shows a fairly constant variance for $z_t$. We will further investigate
this issue in the evaluation of density forecasts using the
probability integral transform (PIT), where we see that the conditional
variance models are indeed capturing the changes in volatility better
and thus generate more reliable density forecasts.

As discussed in Section \ref{sec:Intro}, one may argue that
spatiotemporal information among individual wind farms should be
deployed to forecast aggregated wind power. To show that it is indeed
better to forecast the aggregated power as a univariate time series, we
consider a simple multiple time series approach. We obtain the best
linear unbiased predictor (BLUP) of wind power generation at a single
wind farm using observations in the neighborhood, where the predictor
is the best in the sense that it minimizes mean square errors. In other
words, it is simply the kriging predictor which is widely applied in
spatial statistics [\citet{Cressie1993}, \citet{Stein1999}]. It can be
easily extended to deal with spatiotemporal data
[\citet{Gneiting2007c}], and more details could be found in
\citet{Lau2010}. Computing the BLUP relies on the knowledge of the
covariances of the process between different sites. In the context of
spatiotemporal data, we obtain the BLUP by calculating the empirical
covariances among the wind power at different spatial as well as
temporal lags.\footnote{One needs to decide the number of temporal lags
to be included in calculating the BLUP. In our case of
empirical covariances, we find that
including temporal lags within the past
hour is generally the best. Forecast
performances deteriorate when one
considers too many temporal lags.} We then substitute the empirical covariances into the formula of
BLUP. We apply this method and obtain 1, 6, 12 and 24 hours ahead point
forecasts for the power generated at each individual wind farm,
aggregate all power and normalize the result by dividing by 792.355~MW.
We compute the RMSE of these aggregated forecasts, and find that
aggregating individual forecasts cannot beat the performances of our
approaches in Section \ref{sec:Approach}. The results are displayed in
Table \ref{tab:SummaryRMSE}. Of course, one may expect that more
sophisticated spatiotemporal models may be able to outperform our
methods here, but this will be of more interest to individual power
generation instead of aggregated ones as discussed in this paper.

\begin{table}
\caption{Summary of point forecast performances of different approaches
under RMSE. The bold numbers indicate the best approach at that
forecast horizon} \label{tab:SummaryRMSE}
\begin{tabular*}{\textwidth}{@{\extracolsep{\fill}}lcccc@{}}
  \hline
  & \textbf{1 hour} & \textbf{6 hours} & \textbf{12 hours} & \textbf{24 hours}  \\
  \hline
  Persistence forecast & 0.036 & 0.138 & 0.191 & 0.229 \\
  Constant forecast & $0.263$ & $0.263$ & $0.263$ & $0.263$ \\
  Climatology forecast & 0.285 & 0.285 & 0.285 & 0.285 \\
  EWMA conditional density & 0.177 & 0.192 & 0.203 & 0.211 \\
  $\operatorname{ARIMA}(2,1,3)$ & 0.032 & ${0.118}$ & ${0.177}$ & $\bolds{0.207}$ \\
  $\operatorname{ARIMA}(4,1,3)$--$\operatorname{GARCH}(1,1)$ & $\bolds{0.031}$ & $\bolds{0.117}$ & ${0.177}$ & 0.209 \\
  $\operatorname{ETS}(A,N,N|\mathit{EC})$ & 0.032 & 0.126 & 0.193 & 0.230 \\
  $\operatorname{ETS}(A,N,N|\mathit{EC})$--($A,N,N|\mathit{EC}$) & 0.034 & 0.126 & $\bolds{0.176}$ & 0.215 \\[3pt]
  BLUP (Multiple time series approach) & 0.037 & 0.123 & 0.188 & 0.229 \\
  \hline
\end{tabular*}
\end{table}

\subsection{Density forecasts}

For the density forecasts, we use the continuous ranked probability
score (CRPS) to rank the performances. \citet{Gneiting2007a}
discussed the properties of CRPS extensively, showing that it is a
strictly proper score and a lower score always indicates a better
density forecast. CRPS has become one of the popular tools for density
forecast evaluations, especially for ensemble forecasts in meteorology.
We have also analyzed the performances of density forecasts using other
common metrics such as the negative log likelihood (NLL) scores.
However, we advocate the use of CRPS for ranking different approaches
since CRPS is more robust than the NLL scores, while the latter is
always severely affected by a few extreme outliers
[\citet{Gneiting2005}]. One may need to calculate the trimmed mean
of the NLL scores in order to resolve this problem
[\citet{Weigend2000}]. Also, CRPS assesses both the calibration
and the sharpness of the density forecasts, while the NLL scores
assesses sharpness only.

Similar to evaluating point forecasts, we generate $h$-step ahead
density forecasts for each of the 5504 values in the testing set where
$1 \leq h \leq 96$. For each $h$-step ahead density forecast
$f_{t+h|t}$, let $F_{t+h|t}$ be the corresponding cumulative
distribution function. The CRPS is computed as
\begin{equation}
    \mathit{CRPS} = \int_0^1 [F_{t+h|t}(y) - \mathbf{1}(y-y_{t+h})]^2 \,dy,
\end{equation}
where $\mathbf{1}(\cdot)$ is the indicator function which is equal to
one when the argument is positive. Again, the mean CRPS is taken over
the 5504 $h$-step ahead density forecasts in the testing set.

\begin{figure}[b]

\includegraphics{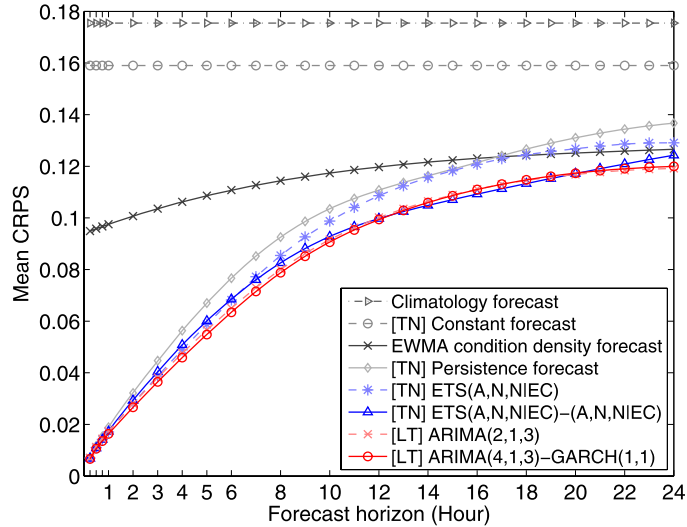}

\caption{Mean continuous ranked probability score (CRPS) of density
forecasts generated by different approaches for forecast horizons from
15 minutes to 24 hours ahead. Rankings are similar to those under MAE
and RMSE in point forecasts.} \label{fig:crps}
\end{figure}

\begin{table}
\caption{Summary of density forecast performances of different
approaches under CRPS. The bold numbers indicate the best approach at
that forecast horizon} \label{tab:SummaryCRPS}
\begin{tabular*}{\textwidth}{@{\extracolsep{\fill}}lcccc@{}}
  \hline
  & \textbf{1 hour} & \textbf{6 hours} & \textbf{12 hours} & \textbf{24 hours}  \\
  \hline
  Persistence forecast & 0.019 & 0.077 & 0.111 & 0.137 \\
  Constant forecast & $0.159$ & $0.159$ & $0.159$ & $0.159$ \\
  Climatology forecast & 0.175 & 0.175 & 0.175 & 0.175 \\
  EWMA conditional density & 0.098 & 0.111 & 0.120 & 0.127 \\
  $\operatorname{ARIMA}(2,1,3)$ & 0.017 & 0.065 & ${0.100}$ & $\bolds{0.119}$ \\
  $\operatorname{ARIMA}(4,1,3)$--$\operatorname{GARCH}(1,1)$ & $\bolds{0.016}$ & $\bolds{0.063}$ & $\bolds{0.099}$ & 0.120 \\
  $\operatorname{ETS}(A,N,N|\mathit{EC})$ & 0.017 & 0.068 & 0.109 & 0.129 \\
  $\operatorname{ETS}(A,N,N|\mathit{EC})$--($A,N,N|\mathit{EC}$) & 0.017 & 0.069 & 0.100 & 0.124 \\
  \hline
\end{tabular*}
\end{table}

\begin{figure}

\includegraphics{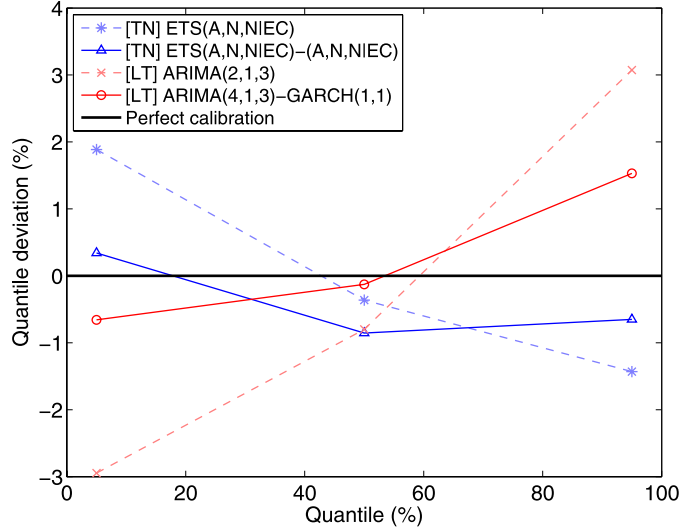}

\caption{We calculate the percentages $P_5$, $P_{50}$ and $P_{95}$ of PIT
values smaller than 0.05, 0.5 and 0.95 respectively, and calculate the
deviations $(P_5-5)$, $(P_{50}-50)$ and $(P_{95}-95)$. The
$\operatorname{ETS}(A,N,N|\mathit{EC}$)--($A,N,N|\mathit{EC}$) method and the
$\operatorname{ARIMA}(4,1,3)$--$\operatorname{GARCH}(1,1)$
model indeed generate better calibrated density forecasts. The overall
calibration of the $\operatorname{ETS}(A,N,N|\mathit{EC}$)--($A,N,N|\mathit{EC}$) method is the best,
indicating that it provides the most reliable descriptions of the
changing volatility over time. Note that a positive slope implies a
ddnsity forecast which is over-conservative, while a negative slope
implies the opposite.} \label{fig:PIT_hor1}
\end{figure}

Figure \ref{fig:crps} shows the performances of density forecasts under
mean CRPS. The rankings are similar to those under MAE and RMSE in
point forecasts. The two ARIMA--GARCH models outperform all other
approaches for all forecast horizons. Table \ref{tab:SummaryCRPS}
summarizes the main results. Again, the performances of the
$\operatorname{ARIMA}(2,1,3$) model are very similar to that of the
$\operatorname{ARIMA}(4,1,3)$--$\operatorname{GARCH}(1,1)$ model and, in contrast, the
$\operatorname{ETS}(A,N,N|\mathit{EC})$--$(A,N,N|\mathit{EC})$ method is
significantly better than the $\operatorname{ETS}(A,N,N|\mathit{EC})$ method.
To investigate the value of including the dynamics of conditional
variances, we consider the probability integral transform (PIT). For
one-step ahead density forecasts $f_{t+1|t}$, the PIT values are given
by
\begin{equation}
    z(y_{t+1}) = \int_0^{y_{t+1}} f_{t+1|t}(y) \,dy.
\end{equation}
\citet{Diebold1998} show that the series of PIT values $z$ are
i.i.d. uniform if $f_{t+1|t}$ coincides with the true underlying density
from which $y_{t+1}$ is generated. For each forecasting approach, we
calculate the percentage of PIT values below the 5th, 50th and
95th quantiles of the $U[0,1]$ distribution, that is, the
percentage of PIT values smaller than 0.05, 0.5 and 0.95 respectively.
We denote them by $P_5, P_{50}$ and $P_{95}$, and calculate the
deviations of the percentages $(P_5-5), (P_{50}-50)$ and $(P_{95}-95)$.
Figure \ref{fig:PIT_hor1} shows the deviations, where we only focus on
the two ETS methods and the two ARIMA--GARCH models. We see that the
$\operatorname{ETS}(A,N,N|\mathit{EC})$--($A,N,N|\mathit{EC}$) method and the
$\operatorname{ARIMA}(4,1,3)$--$\operatorname{GARCH}(1,1)$ model indeed generate density forecasts which
are better calibrated. In particular, the overall calibration of the
$\operatorname{ETS}(A,N,N|\mathit{EC})$--($A,N,N|\mathit{EC}$) method is the
best, indicating that it provides the most reliable descriptions of the
changing volatility over time. Note that a positive slope in Figure
\ref{fig:PIT_hor1} implies a density forecast which is
over-conservative and has a large spread, while a negative slope
implies the opposite. Thus, for one-step ahead forecasts, the
ARIMA--GARCH models are over-conservative, while the ETS methods are
over-confident.

\begin{figure}

\includegraphics{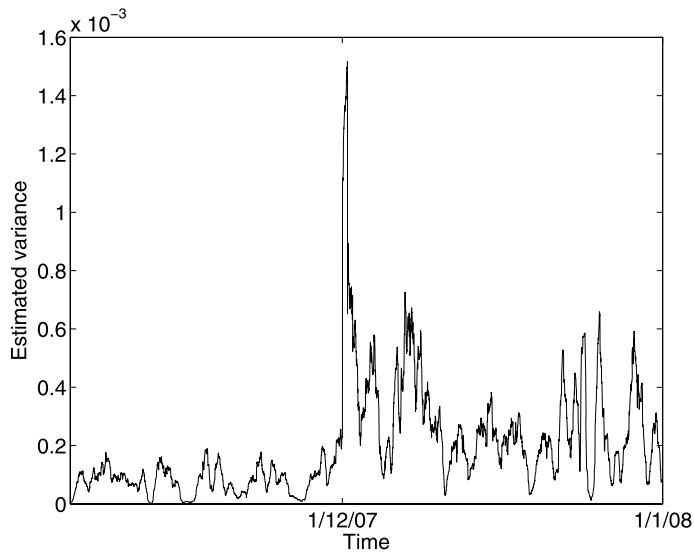}

\caption{Estimated variance of data in the testing set using the
persistence forecast $\hat{s}_{\varepsilon;t+1|t}^2$ in
(\protect\ref{eq:persistence}), which essentially gives the 12-hour moving
average of realized variance. Clearly, the variance changes with time
and the largest values mostly occur in early December.}
\label{fig:EstimatedVariance}
\end{figure}

\begin{figure}

\includegraphics{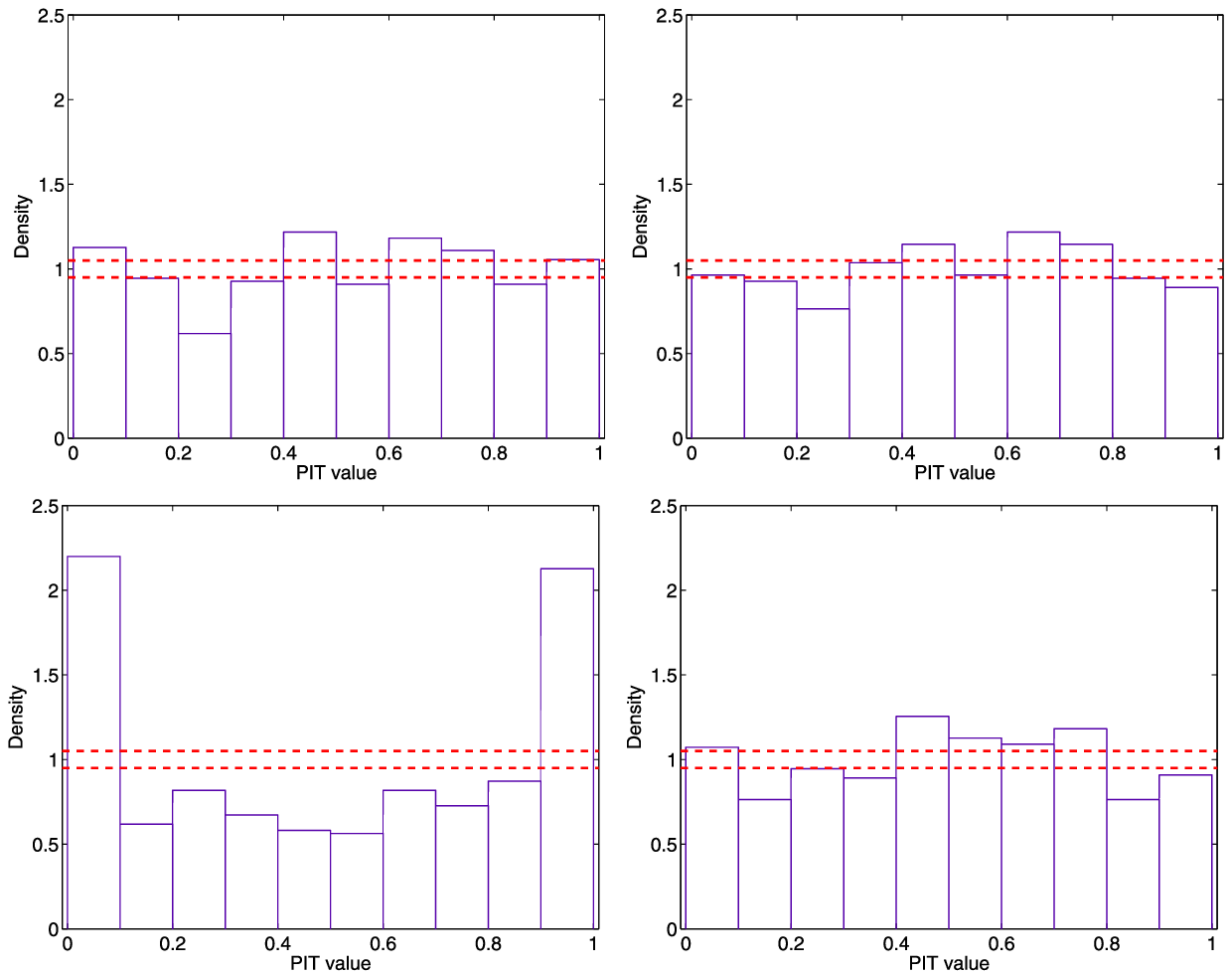}

\caption{Histograms of PIT values conditioned on the largest 10\% of
estimated variance, where the one-step ahead density forecasts are
generated using the $\operatorname{ARIMA}(2,1,3)$ model (top left), the
$\operatorname{ARIMA}(4,1,3)$--$\operatorname{GARCH}(1,1)$ model (top right),
the $\operatorname{ETS}(A,N,N|\mathit{EC}$) method
(bottom left) and the $\operatorname{ETS}(A,N,N|\mathit{EC}$)--($A,N,N|\mathit{EC}$) method (bottom
right). The dotted lines correspond to 2 standard deviations from the
uniform density.} \label{fig:PIT_largeV}
\end{figure}

Figure \ref{fig:PIT_hor1} only reflects information on the marginal
distributions of the PIT values. \citet{Stein2009} suggests that
it is also valuable to evaluate the distributions conditioned on
volatile periods. It is particularly important to capture the variance
dynamics during times of large volatilities, since for most of the
times one does not want to underestimate the risk by proposing an
over-confident density forecast. Underestimating large risks usually
leads to a more disastrous outcome than overestimating small risks.
Following \citet{Stein2009}, we compare the ability of the
approaches in capturing volatility dynamics during the largest 10\% of
variance. To estimate the variance of the data in the testing set, we
directly adopt the persistence forecast $\hat{s}_{\varepsilon;t+1|t}^2$
in (\ref{eq:persistence}), which essentially gives the 12-hour moving
average of realized variance. Figure \ref{fig:EstimatedVariance} shows
the changing variance, where the largest values mostly occur in early
December. The times corresponding to the largest 10\% of variance are
selected and we compare the distribution of $z(y_{t+1})$ at those
times. The PIT diagrams are shown in Figure \ref{fig:PIT_largeV}. It
demonstrates that the ARIMA--GARCH model indeed gives better calibrated
one-step ahead density forecasts than the ARIMA model during volatile
periods. The differences between the two ETS methods are even more
significant, where the $\operatorname{ETS}(A,N,N|\mathit{EC})$ method gives
over-confident density forecasts that underestimate the spread.

\section{Conclusions and discussions}\label{sec:Conclusion}

In this paper we study two approaches for generating multi-step density
forecasts for bounded non-Gaussian data, and we apply our methods to
forecast wind power generation in Ireland. In the first approach, we
demonstrate that the logistic transformation is a good method to
normalize wind power data which are otherwise highly non-Gaussian and
nonstationary. We fit ARIMA--GARCH models with Gaussian innovations for
the logistic transformed data, and out-of-sample forecast evaluations
show that they generate both superior point and density forecasts for
all horizons from 15 minutes up to 24 hours ahead. A second approach is
to assume that the $h$-step ahead conditional densities are described
by a parametric function $D$ with a location parameter $\hat{\ell}$ and
scale parameter $\hat{s}^2$, namely, the conditional mean and the
conditional variance of $y_t$ that are generated by an appropriate
Gaussian model $G$. Results show that choosing $D$ as the truncated
normal distribution is appropriate for aggregated wind power data, and
in this case $\hat{\ell}$ and $\hat{s}^2$ are the mean and variance of
the original normal distribution respectively. We apply exponential
smoothing methods to generate $h$-step ahead forecasts for the location
and scale parameters. Since the underlying models of the exponential
smoothing methods are Gaussian, we are able to obtain multi-step
forecasts by simple iterations and generate forecast densities as
truncated normal distributions.

Although the approach using exponential smoothing methods with
truncated normal distributions cannot beat the approach considering
logistic transformed data, they are still a useful alternative to
produce good density forecasts due to several reasons. First, forecast
performances of the exponential smoothing methods are more robust under
different lengths of training data, especially when the size of the
training set is relatively small and the estimation of the ARIMA--GARCH
models may not be reliable to extrapolate into the testing set. This
has been demonstrated in our data, where we take 40\% of the data as
the training set and the remaining as the testing set. In such a case,
the $\operatorname{ETS}(A,N,N|\mathit{EC})$--($A,N,N|\mathit{EC}$) method performs
better than the approach with logistic transformed data [Lau (\citeyear{Lau2010})]. Second, in the
first approach using ARIMA--GARCH models, we have to select the best
model using BIC whenever we consider an updated training set. This is
not necessary for the exponential smoothing methods. Third, an
advantage of forecasting by exponential smoothing methods is that it is
computationally more efficient to calculate point forecasts due to the
closed form of density function that we have chosen, namely, the
truncated normal distribution $D$. On the other hand, in the first
approach, we have to transform the Gaussian densities and calculate the
expected value of the transformed densities by numerical integrations,
which require much more computational power. The second and third
points are critical since, in practice, many forecasting problems
require frequent online updating. Finally, the second approach allows
us to choose a parametric function $D$ for the forecast densities,
which gives us more flexibility and one may generate improved density
forecasts by testing various possible choices of $D$. This advantage is
particularly important when there are no obvious transformations to
normalize the data, and when there is evidence that supports simple
parametric forecast densities.

In summary, we have developed a general approach of generating
multi-step density forecasts for non-Gaussian data. In particular, we
have applied our approaches to generate multi-step density forecasts
for aggregated wind power data, which would be economically valuable to
power companies, national grids and wind farm operators. It would be
interesting and challenging to propose modified methods based on our
current approaches, so that reliable density forecasts for individual
wind power generation could be obtained. Individual wind power time
series are interesting since they are highly nonlinear. Sudden jumps
from maximum capacity to zero may occur due to gusts of winds, and
there may be long chains of zero values because of low wind speeds or
maintenance of turbines. Characteristics of individual wind power
densities include a positive probability mass at zero as well as a
highly right-skewed distribution, and it would be challenging to
generate multi-step density forecasts for individual wind power data.
Another important area of future research is to develop spatiotemporal
models to generate density forecasts for a portfolio of wind farms at
different locations. Recent developments in this area include
\citet{Hering2009}. Some possible approaches include the
process-convolution method developed and studied by
\citet{Higdon1998}, which has been applied to the modeling of
ocean temperatures and ozone concentrations. Another possible approach
is the use of latent Gaussian processes. Those approaches have been
studied by \citet{Sanso1999} who consider the power truncated
normal (PTN) model, and by \citet{Berrocal2008} who consider a
modified version of the PTN model called the two-stage model.
Spatiotemporal models will be important to wind farm investors to
identify potential sites for new farms. It would also be of great
importance to the national grid systems where a large portfolio of wind
farms are connected, and sophisticated spatiotemporal models may be
constructed to improve density forecasts for aggregated wind power by
exploring the correlations of power generations between neighboring
wind farms.

\section*{Acknowledgments}
The wind power generation data in Ireland was
kindly provided by Eirgrid plc. The authors thank
Pierre Pinson, James Taylor, Max Little, the referees and the associate
editors for insightful comments and suggestions.

\printaddresses

\end{document}